# Black phosphorus van der Waals heterostructures light emitting diodes for mid-infrared silicon photonics

**Tian-Yun Chang[1], Yueyang Chen[2], De-In Luo[1], Jia-Xin Li[1], Po-Liang Chen[1], Seokhyeong Lee[2], Zhuoran Fang[2], Wei-Qing Li[1], Ya-Yun Zhang[1], Mo Li[2,3], Arka Majumdar[2,3,\*], Chang-Hua Liu[1,4,\*]**

1. Institute of Photonics Technologies, National Tsing Hua University, Hsinchu 30013, Taiwan

2. Department of Electrical and Computer Engineering, University of Washington, Seattle, Washington 98195, United States

3. Department of Physics, University of Washington, Seattle, Washington 98195, United States

4. Department of Electrical Engineering, National Tsing Hua University, Hsinchu 30013, Taiwan

**Light-emitting diodes (LEDs) based on III–V/II–VI materials have delivered a compelling performance in the mid-infrared (mid-IR) region, which enabled wide-ranging applications in sensing, including environmental monitoring, defense, and medical diagnostics[1-3]. Continued efforts are underway to realize on-chip sensors via heterogeneous integration of mid-IR emitters on a silicon photonic chip[4-6]. But the uptake of such approach is limited by the high costs and interfacial strains, associated with the processes of heterogeneous integrations. Here, the black phosphorus (BP)-based van der Waals (vdW) heterostructures are exploited as room temperature LEDs. The demonstrated devices can emit linearly polarized light ($\rho$=75 %), and their spectra cover the technologically important mid-IR atmospheric window (3-4 μm). Additionally, the BP LEDs exhibit fast modulation speed as well as exceptional operation stability, and its peak extrinsic quantum efficiency (QE~0.9%) is comparable to the III–V/II–VI mid-IR LEDs[3,7]. By leveraging the integrability of vdW heterostructures, we further demonstrate a silicon photonic waveguide-integrated BP LED. The reported hybrid platform holds great promises for mid-IR silicon photonics.**

**Main text:**

The development of light sources compatible with silicon platform is essential for silicon photonic technologies. By far, exploiting germanium or III–V/II–VI semiconductors for on-chip light sources has long been pursued, and multiple research works have successfully demonstrated their spectral coverage can extend from the near-IR to mid-IR region[1-3]. This potentially expands the applications of silicon photonics from traditional telecommunication applications to lab-on-chip spectroscopic sensing, as many chemical species have strong absorption fingerprints in the mid-IR band[4-6]. Despite continued advancements, the current on-chip sources are realized via wafer bonding or direct heteroepitaxial growth of germanium and III–V/II–VI on the silicon chip. Such an approach would result in increased fabrication costs and complex device structures. Furthermore, these heterogeneous integration processes would cause the interfacial strains, originated from the lattice- and thermal-mismatch, and consequently, lead to luminescence quenching of on-chip emitters[4-6].

To that end, utilizing van der Waals (vdW) materials have recently emerged as a promising alternative approach. One of the key reasons is that vdW family members cover a wide range of optical bandgaps, offering the opportunities of developing vdW-based optoelectronics, operating at different spectral regions[8-10]. In addition, due to the weak vdW bonding of these materials, diverse vdW materials can be vertically stacked to form heterostructures. Further, these materials and their heterostructures can be assembled with integrated photonic structures, such as cavities or waveguides, without stringent requirement on lattice matching[11-15]. However, the vdW-based LEDs reported thus far generally exploit the family of transition metal dichalcogenides as the emissive layers, which can operate only in the visible to near-infrared region[8,10,12,13], and their applications on the longer spectral regions are largely unexplored. In this paper, the vdW heterostructures with BP light emitting layers are explored as LEDs in the mid-IR wavelength range. By leveraging the direct narrow gap (~0.3 eV) and anisotropic optoelectronic properties of BP[16-18], we show such device can emit linearly polarized mid-IR light with the extrinsic QE comparable to the III–V/II–VI devices. Via integrating with a silicon waveguide, we further demonstrate that the emission of BP-based LED can evanescently couple and propagate through the waveguide, showing its potential applications in on-chip mid-IR light source.

Figures 1a-b show the optical micrograph and schematic of hBN-encapsulated LED, composed of the 87-nm-thick BP sandwiched by two graphite electrodes (see Methods). The bulk BP, used here as the light emitting layer, exhibits three characteristic Raman peaks (Fig. 1c). The top ($Gr_T$) and bottom ($Gr_B$) graphite electrodes are semi-transparent in the mid-IR region[19] and capable of injecting high density of current into BP. To probe the basic electrical properties of heterostructures, the bias voltage ($V_b$) was applied across two graphite contacts with the $Gr_T$ being grounded. As shown in Fig. 1d, the device exhibits nonlinear and asymmetric *I-V* characteristics, which reflects the dissimilarities of two Schottky barriers at BP/Gr interfaces[20]. The detailed information on the band bending of two barriers can be resolved using the scanning photocurrent microscopy (see Supplementary Section 1). In addition, it is notable that the graphite contacts can inject up to 6.7 mA current into ~1200 $\mu m^2$ BP flake. Such high current density ($5.58 \times 10^5$ mA/$cm^2$) injection could potentially lead to the detectable electroluminescence signal.

We built the experimental setup as schematically illustrated in Fig. 2a (also see Methods) to characterize the luminescence of vdW heterostructures. As an initial experiment, the bias voltage ($V_b$=6 V) was applied on the LED, mounted on the x-y scanning stage. The generated electroluminescence signal was mechanically chopped and detected by a liquid nitrogen cooled indium antimonide (InSb) detector as well as the lock-in technique (Fig. 2a, green beam path). This allows us to resolve the spatial distribution of electroluminescence (Fig. 2b). To gain further insight, we also conducted the scanning reflection experiment, which is helpful to identify the position of the objective focal point on the measured device. The excitation wavelength is selected at 3.5 µm, as BP is highly absorbing in this spectral region, leading to the better reflectivity contrast of image (Fig. 2c). Notably, by comparing these two scanning images, it is clear the emission originates from the heterostructures and its peaks from the region, where $Gr_T$ and $Gr_B$ are overlapped with each other.

We next analyzed the spectral characteristic of BP emitter via sending the electroluminescence signals to a monochromator, as shown in the red beam path in Fig. 2a (see Methods). Figure 2d presents the electroluminescence spectra for different bias voltage $V_b$. The results show that all spectra have the same peak wavelength ~3.64 µm, which is also in good agreement with our photoluminescence measurement of BP (Fig. 2e)[21]. Such a feature indicates that the measured electroluminescence is originated from the recombination of electrons and holes, injected from graphite contacts, rather than the thermal black body radiation, as the peak wavelength is closed to the band gap of bulk BP. However, it is notable that the spectral width of electroluminescence at $V_b = 1$ V is ~0.585 µm (calculated by measuring the full width at half maximum, FWHM), which is wider than the linewidth of photoluminescence (FWHM ~0.423 µm). In addition, when tuning the $V_b$ to 8 V, the FWHM of electroluminescence further increases to ~0.65 µm. The observed spectral broadenings suggest that the large vertical field (as high as 0.9 MV/cm at $V_b = 8$ V) and the high current injection density could redistribute the carriers into higher energy states in the momentum space of BP and also could enhance the trionic emission. A similar phenomena have been reported in literature for other vdW materials in the past[12,13,22,23].

After characterizing the emission spectrum, we calibrated the output power of BP mid-IR LED, operated at various bias voltages (see Supplementary Section 2). The result shown in Fig. 2f reveals that bipolar charge injection into the emissive BP layer can occur at very low bias voltage ($V_b$~ 0.45 V). The BP with light emitting area ~1200 $\mu m^2$ can produce output power approaching 10 $\mu W$, and can also exhibit excellent operation stability over the periods of at least 9 hours (see Supplementary Section 3). These features highlight the usefulness of our BP LED for practical applications. The extrinsic QE ($\eta$) of our BP LED can be quantified by [11]:

$$\eta = \frac{2qN}{I}$$

where $q$ is the electron charge, $N$ the number of photons emitted into the free space per second, and $I$ the current injected into BP LED (see Supplementary Section 2). As shown in Fig. 2g, the extrinsic QE can approach 0.9% at $V_b = 3.5$ V. This performance is comparable to other III-V mid-IR LEDs[3,7]. By further increasing the bias voltage to 8 V, the efficiency slightly decreases to 0.83%, showing the small reduction of efficiency with increasing current. Notably, this low-efficiency droop performance can be observed in multiple BP LEDs we fabricated (see Supplementary Section 4), and such feature is difficult to achieve for many vertical heterostructures LEDs, based on the vdW, organic or III-V emissive materials. This is because the high vertical field and high current density generally cause the charge carriers passing through the contacts without forming excitons and increase the nonradiative Auger process[2,22,24].

To investigate the polarization properties of the BP emitter, a linear polarizer was added in the light collection path (green beam path, Fig. 2a). Figure 3a presents a colour map of the polarization angle-dependent photoluminescence amplitude of BP, as a function of excitation power (λ=2.5 µm). The calculated degree of polarization factors ($\rho$) are ~80% (Fig. 3b), associated with the property of anisotropic optical transition near the band edge of BP[16-18,20,21]. For comparison, we also characterized the emission anisotropy of electroluminescence of the BP LED, under different bias voltages (Fig. 3c). The measured electroluminescence amplitude shows the same polarization angle-dependent variations. The calculated polarization factors of electroluminescence (Fig. 3d) are closed to the photoluminescence results, indicating the possibility of applying BP LEDs as highly polarized mid-IR sources.

Next, we characterized the frequency response of the BP LED via electrically turning it on and off at different modulation frequencies and monitoring the change of photovoltages, detected by the lock-in technique. For this experiment, the detector used to collect electroluminescence signals (Fig. 2a, green beam path) was replaced with a room temperature Indium Arsenide Antimonide (InAsSb) detector due to its faster operation speed (9 MHz bandwidth, Thorlabs PDA07P2). We note that while the InAsSb detector has a mediocre photoresponsivity (~4 mA/W), the frequency dependence of electroluminescence can still be successfully resolved (Fig. 3e). We measured the 3 dB roll-off frequency as ~2 MHz, limited by the bandwidth of the lock-in amplifier. The ultimate speed limit of the reported device is estimated above ~700 MHz [20], related to the lifetime of exciton emission in BP and RC time constant of device (see Supplementary Section 5). These demonstrations clearly show the opportunities of using the reported BP LED for the free space sensing or spectroscopic applications without need of external modulators.

Finally, we explored the potential of using our BP heterostructure LEDs for mid-IR integrated silicon photonics. Figure 4a shows a scanning electron micrograph (SEM) of the hybrid device: the BP LED is integrated onto a mid-IR silicon waveguide with two grating couplers at both ends. The details of the design, fabrication and simulation are described in Methods and Supplementary Sections 6. When applying the bias voltage across two graphite contacts, three bright spots can be observed from the scanning electroluminescence map (Fig. 4b), distinct from the scanning result shown in Fig. 2b. Moreover, by comparing the scanning electroluminescence with the reflection maps (Fig. 4c), it is clear the central brighter spot corresponds to the position of BP LED, while the two extra emission spots are located at the grating couplers. These evidently confirm that the mid-IR electroluminescence of BP LED can be evanescently coupled to the silicon photonic waveguide. The coupled light then propagates through the waveguide to the coupler, which redirects the light out to the free space. We note that the coupling of BP emission into the silicon waveguide can also be validated via three-dimensional finite-difference time domain (FDTD) simulations (Fig. 4d). However, the measured emission intensity from the right (left) grating coupler is 10% (6%) with respect to the central BP emissive region, which are lower than the simulation result (40%, see Supplementary Sections 6). This suggests the performance of our waveguide-integrated BP LED can be improved by further optimizing the silicon photonics fabrications and minimizing the residue remaining on the waveguide due to the transfer of vdW heterostructures.

In summary, we report a new approach to realize room temperature mid-IR LEDs that exhibit exceptional linear degree of polarization, long-term stability, high extrinsic QE and fast modulation speed. Further engineering of the reported heterostructures will open up new technological opportunities. For example, the emission wavelength could be tuned via applying the mechanical strain on the BP heterostructures[25], and emission bandwidth could be expanded into multiple spectral regions by implementing different band gaps of vdW materials into the BP-based heterostructures[8,9]. Furthermore, we demonstrate integration of the BP LED with a silicon waveguide. Considering the recent advancements in scalable synthesis of vdW materials[26], we expect that the BP LEDs can be potentially integrated in large-scale photonic integrated circuits, offering tremendous opportunities for mid-IR silicon photonics applications.

## Methods:

### Fabrication of BP LEDs

The vdW materials present in this work were all prepared by using mechanical exfoliation method, and their thicknesses were identified by atomic force microscopy (AFM). To create the vdW heterostructures, a dry transfer technique was utilized to assemble different materials in the vertical direction[27]. The vertically-stacked hBN/$Gr_T$/BP/$Gr_B$ heterostructures was subsequently transferred onto a $SiO_2$ substrate with prepatterned Cr/Au (5/40 nm) metal electrodes, and during this transfer process, the $Gr_T$ and $Gr_B$ were aligned to attach the two separate metal electrodes. The fabricated BP LEDs were then loaded into a vacuum chamber ($10^{-4}$ torr) to characterize their optoelectronic properties at room temperature.

### Measurement of emission powers and spectra of BP

The green beam path shown in Fig. 2a was used to measure the emission powers of the BP LED. For this experiment, the bias voltage was applied on two metal electrodes. The emitted electroluminescence was mechanically chopped and focused onto a liquid nitrogen cooled InSb photodetector (Infrared Associates, IS-2.0). The generated photovoltage was then detected by the lock-in amplifier (Stanford Research Systems, SR865). The detected photovoltage can be converted into the optical power by characterizing the photoresponsivity of InSb photodetector (see Supplementary Section 2).

To analyze the spectra of electroluminescence, the light emitted by the BP LED was also mechanically chopped and passed through the red beam path shown in Fig. 2a. But this beam path has an additional grating mounted on the rotation stage. The relation between the rotated angle and wavelength detected by the InSb detector was calibrated by coupling the laser, generated by a wavelength tunable mid-IR optical parametric oscillator (OPO), into the red beam path. For the photoluminescene experiments, a laser beam ($\lambda$=2.5 µm) was coupled into the experimental setup to excitate the BP flake. The generated photoluminescence passed thorugh a band pass optical filter to get rid of the excitation beam reflected from the BP flake, and its spectrum was analyzed using the setup along the red beam path, the same as the electroluminescence measurements.

### Fabrication of the waveguide-integrated BP LED

The silicon waveguide and grating couplers were fabricated on an SOI chip that consists of a 600nm of crystalline silicon, 2 µm of oxide and 500 µm silicon substrate. To create these silicon photonics devices, we spun coat ZEP520A resist on the SOI chip and created the patterns using the ebeam lithography (JEOL JBX6300FX). The patterns were then transferred to the silicon using an inductive-coupled plasma etching in SF6/C4F8 chemistry. The ZEP520A resist was then stripped off by piranha solution. After fabricating the waveguides and grating couplers, two metal contacts were further defined by ebeam overlay, following with atomic layer deposition (ALD), ebeam evaporation and liftoff. The ALD process deposited 10 nm alumina on the silicon, and the ebeam evaporation process deposited Cr/Au (5/50 nm) as metal contacts. To create the waveguide-integrated BP LED, we exploited the dry transfer technique to integrate the hBN/$Gr_T$/BP/$Gr_B$ heterostructures onto the silicon waveguide, with the $Gr_T$ and $Gr_B$ attached to two separate metal contacts, and the armchair direction of the BP flake is perpendicular to the silicon waveguide. The fabricated hybrid device was performed at room temperature under a vacuum of $10^{-4}$ torr.

**Author contributions:**

C.-H.L. and A.M. supervised the project. J.-X.L. and W.-Q.L. exfoliated and characterized the vdW materials. D.-I.L. and Y.-Y.Z. fabricated the vdW heterostructures. Y.C. fabricated the silicon photonics, assisted by S.L. Z.F. and M.L. T.-Y.C. performed the measurements, assisted by P.-L.C and C.-H.L. All authors contributed to the discussion of the data in the manuscript and Supplementary Information.

**Acknowledgements:**

The authors thank Prof. Chen-Bin Huang for sharing electronic instruments. This work was supported by the Ministry of Science and Technology of Taiwan under grant 107-2112-M-007-002-MY3, NSF-1845009 and NSF-ECCS-1708579. A.M. also acknowledges support from Sloan Foundation.

**Competing interests**
The authors declare no competing financial interests.

**Figures:**

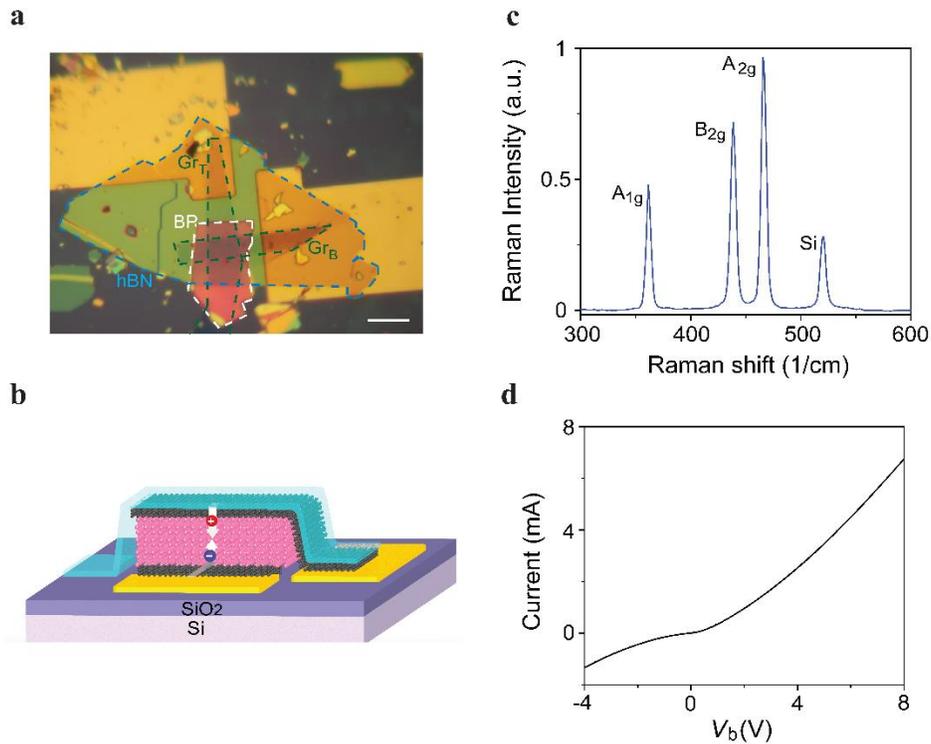

**Figure 1 BP-based vdW heterostructures. a.** Optical microscope image of the BP-based vdW heterostructures. The regions of Gr, BP and hBN are defined by green, white and blue dashed lines respectively. Scale bar, 20 µm. **b.** Schematic of the device configuration, composed of vertically stacked hBN/$Gr_T$/BP/$Gr_B$ heterostructures. The injected electrons and holes from two Gr contacts would recombine in the BP flake, leading to electroluminescence. **c.** Raman spectrum of the used BP flake, showing three characterisitic Raman peaks. The spectrum was collected with λ=532 nm excitation. **d.** *I-V* characterisitic of the BP-based heterostructures.

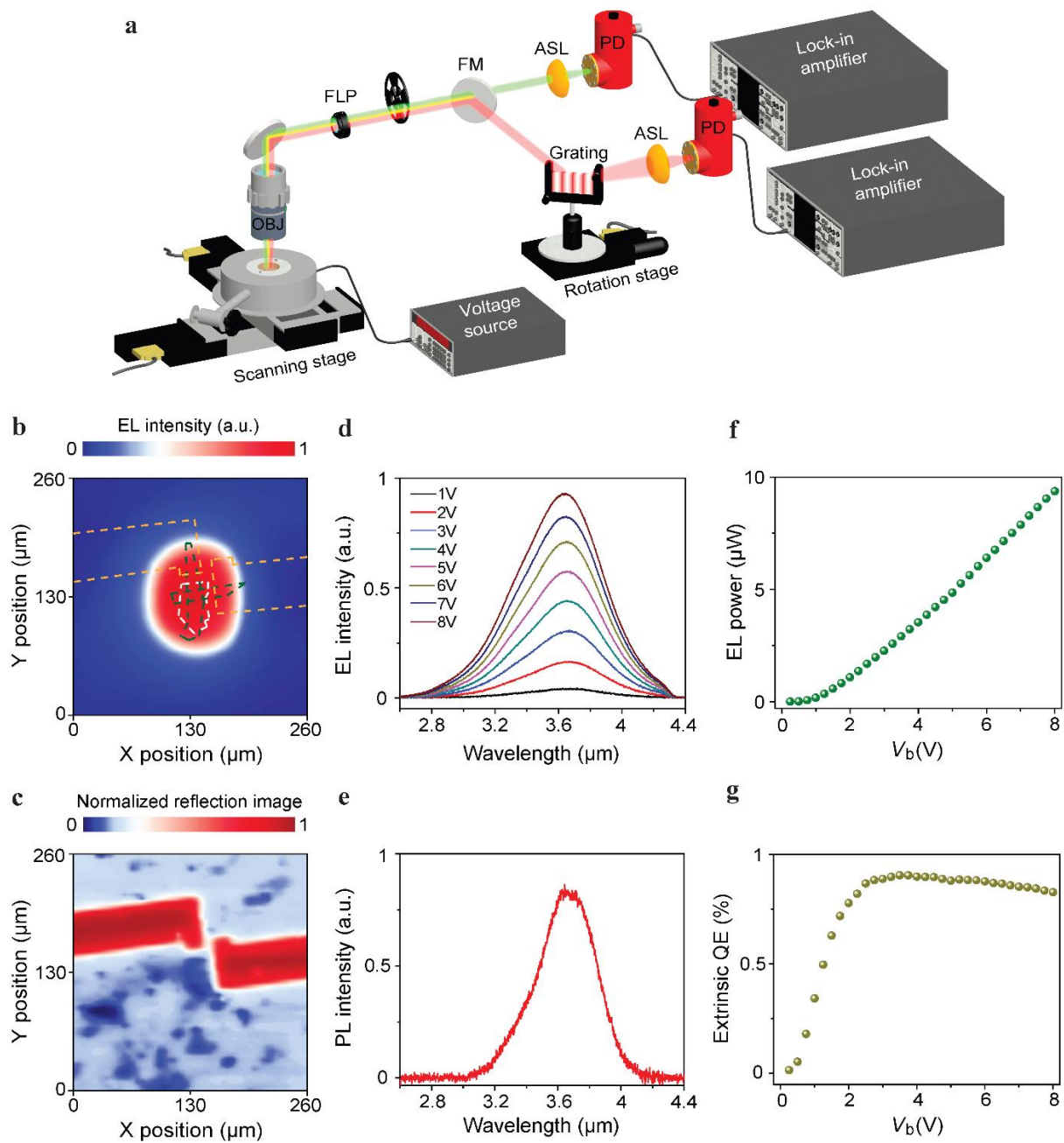

**Figure 2 Optoelectronic characterizations of the BP-based vdW heterostructures. a.** Schematic of the experimental setup that can perform the electrical transport, as well as the scanning photocurrent, electroluminescence, and photolumiescence measurements. The green (red) beam path is used to measure the emission power (spectra) of the BP LED. PD, photodetector; FM, flippable mirror; FLP, flippable linear polarizer; ASL, aspheric lens; OBJ, objective (N.A. = 0.67). **b,c,** The spatially resolved two-dimensional (b) electroluminescence map, and (c) reflection map. In panel (b), the regions of Gr, BP and metal contacts are defined by green, white and orange dashed lines respectively. The results were obtained, as the device shown in Fig. 1a was biased at 6 V. **d.** Bias dependence of electroluminescence spectra, emitted by the BP LED. **e.**

Photolumiescence spectrum measured from the used BP flake. The excitation wavelength is 2.5 µm and excitation power is 500 µW. **f.** Device emission powers at various bias voltages. **g.** Extrinsic QE of the BP LED, operated at various bias voltages.

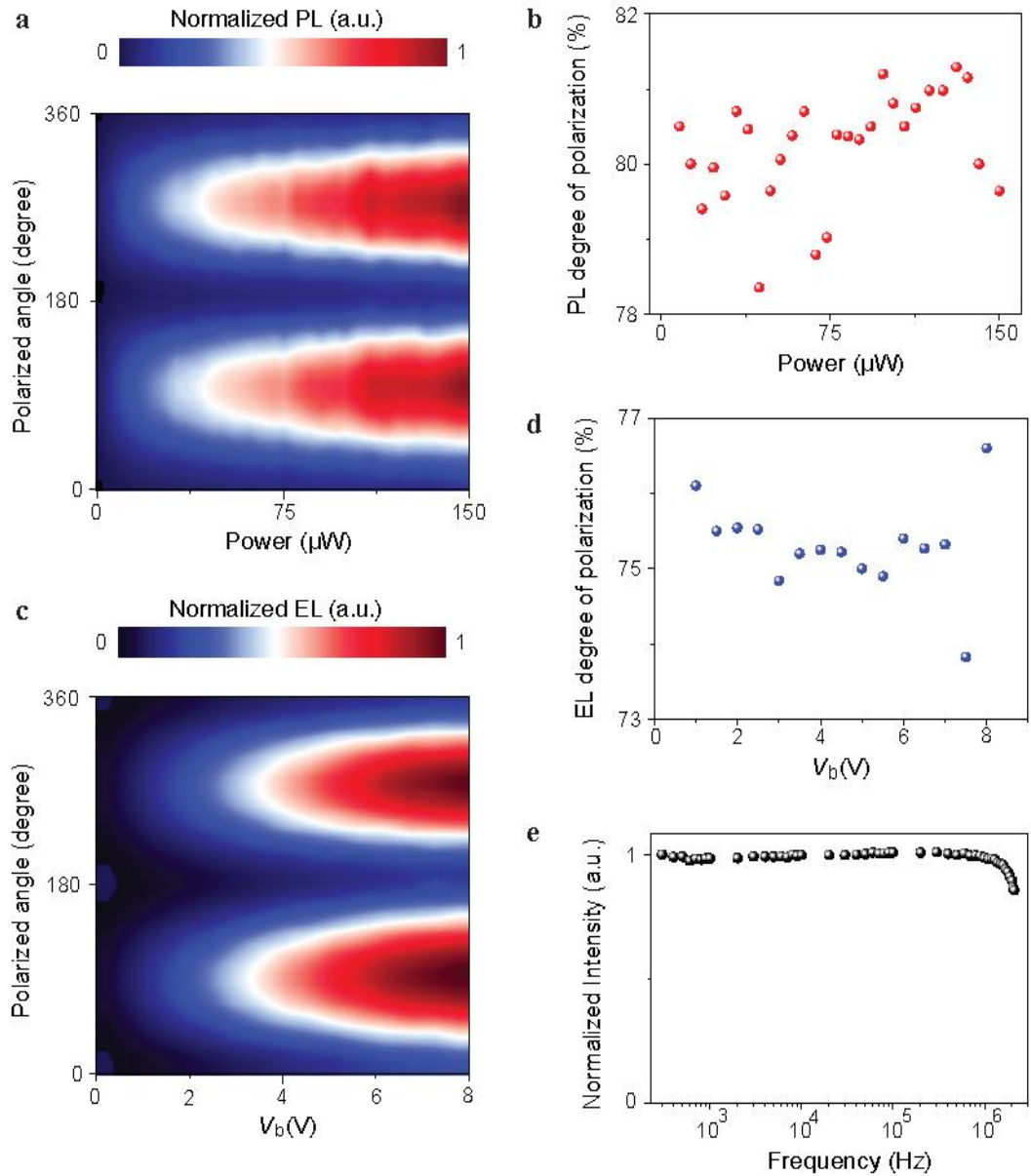

**Figure 3 Polarization-resolved and frequency-dependent electroluminescence. a.** Linear polarization-dependent mapping of BP photolumiescence intensity ($I_{PL}$) under various excitation powers ($\lambda=2.5$ μm). **b.** Power-dependent degree of polarization of photoluminescence, which is quantified by the formula $\frac{I_{PL,max} - I_{PL,min}}{I_{PL,max} + I_{PL,min}}$. The calculations are based on the results shown in panel (a). **c.** Linear polarization-dependent mapping of BP electroluminescence intensity ($I_{EL}$) under various bias voltages. **d.** Bias-dependent degree of polarization of electroluminescence, which is quantified by the formula $\frac{I_{EL,max} - I_{EL,min}}{I_{EL,max} + I_{EL,min}}$. The calculations are based on the results shown in panel (c). **e.** Frequency response of the BP LED. For this experiment, voltage pulses generated from a function generator not only modulated the electroluminescence of BP LED but also served as a reference signal of lock-in amplifier. The modulated electroluminescence signal passed along the green beam path shown in Fig. 2a and was then detected by the InAsSb detector and lock-in amplifier. By tuning the on-off frequency of voltage pulses, the variation of photovoltage measured by the lock-in amplifier was recorded. The measured photovoltage is normalized with respect to the peak intensity.

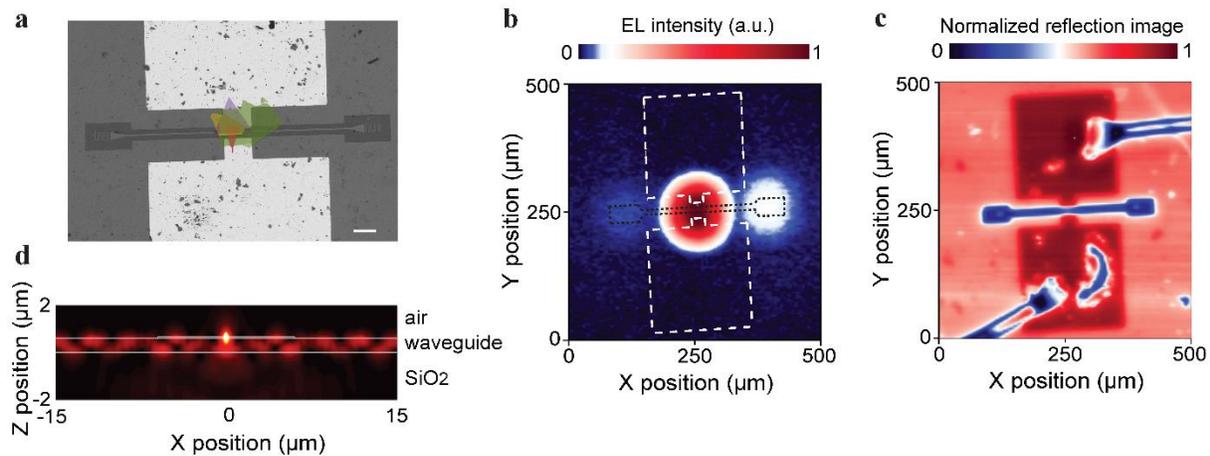

**Figure 4 The waveguide-integrated BP LED. a.** A false-colour SEM image of the waveguide-integrated BP LED. The emitter is composed of the vertically stacked hBN (green)/Gr$_T$ (red)/BP (yellow)/Gr$_B$ (purple) heterostructures. The used BP flake is 65-nm-thick. Scale bar, 50 µm. **b,c,** The spatially resolved two-dimensional (b) electroluminescence map, and (c) reflection map (illumination wavelength: 3.5 µm). The electroluminescence results were obtained, as the device shown in panel (a) was biased at 3 V. In panel (b), the regions of metal electrodes are outlined with white dashed lines, and the regions of waveguide and grating couplers are outlined with black dashed lines. **d.** An FDTD simulation of the cross-section of electric field intensity distribution. In this sumulation, a BP flake is placed on top and at the center of a 600-nm-thick silicon waveguide, and there is a 2-µm-thick silicon oxide layer beneath the silicon waveguide. The thickness and lateral size of BP flake are set to 65 nm and 50 µm respectively. The refractive index of BP is assumed to be (n,k) = (4.4,0.2) [28]. A dipole source with orientation along the y-axis is placed at the center of BP layer.